\documentclass[aps,prb,twocolumn,superscriptaddress]{revtex4-1}

\usepackage[usenames,dvipsnames]{xcolor}
\usepackage[caption=false]{subfig} 

\usepackage{csquotes} 
\usepackage{upgreek} 
\usepackage[version=3]{mhchem} 

\usepackage{tikz}
\usepgflibrary{plotmarks}

\usepackage[pdftex,colorlinks,bookmarks = true]{hyperref} 
\usepackage[capitalize]{cleveref} 

\usepackage{amsmath}
\usepackage{amsfonts}
\usepackage{amssymb}
\usepackage{bm} 
\usepackage{empheq} 
\usepackage{mathtools} 

\usepackage{siunitx} 
\usepackage{braket} 

\let\vaccent=\v 
\renewcommand{\v}[1]{\bm{#1}} 

\newcommand{\inv}{^{-1}} 
\newcommand{\e}[1]{\, \mathrm{e}^{#1}} 
\renewcommand{\i}{\mathrm{i}} 
\newcommand{\cpi}{\uppi} 
\providecommand*{\groupU}{\mathrm{U}(1)} 
\providecommand*{\groupZ}{\mathbb{Z}_2} 
\providecommand*{\groupUZ}{\groupU \times \groupZ} 

\DeclareMathOperator{\Tr}{Tr} 

\makeatletter 
\providecommand*{\diff}
	  {\@ifnextchar^{\DIfF}{\DIfF^{}}}
  \def\DIfF^#1{%
	  \mathop{\mathrm{\mathstrut d}}%
		  \nolimits^{#1}\gobblespace}
  \def\gobblespace{%
	  \futurelet\diffarg\opspace}
  \def\opspace{%
	  \let\DiffSpace\!%
	  \ifx\diffarg(%
		  \let\DiffSpace\relax
	  \else
		  \ifx\diffarg[%
			  \let\DiffSpace\relax
		  \else
			  \ifx\diffarg\{%
				  \let\DiffSpace\relax
			  \fi\fi\fi\DiffSpace}
\makeatother
\providecommand*{\pderiv}[3][]{\frac{\partial^{#1}#2}{\partial #3^{#1}}} 
\providecommand*{\fderiv}[3][]{\frac{\delta^{#1}#2}{\delta #3^{#1}}} 

\usepackage{soul} 
  \soulregister\enquote7
  \soulregister\cref7
  \soulregister\cite7

\begin{document}

\title{Fluctuation effects in phase-frustrated multiband superconductors}

\author{Troels Arnfred Bojesen}
\email{troels.bojesen@riken.jp}
\affiliation{Quantum Matter Theory Research Team, RIKEN Center for Emergent Matter Science, RIKEN, 2-1 Hirosawa, Wako, Saitama 351-0198, Japan}

\author{Asle Sudb\o{}}
\email{asle.sudbo@ntnu.no}
\affiliation{Department of Physics, Norwegian University of Science and Technology, NO-7491 Trondheim, Norway}

\date{\today}

\begin{abstract}
We compare the phase-diagrams of an effective theory of a three-dimensional multi-band superconductor obtained within standard and cluster mean-field theories, and in large-scale Monte Carlo simulations. In three dimensions, mean field theory fails in locating correctly the positions of the phase transitions, as well as the character of the transitions between the different states. A cluster mean-field calculations taking into account order-parameter fluctuations in a local environment improves the results considerably for the case of extreme type-II superconductors where gauge-field fluctuations are negligible. The large fluctuations in the multi-component superconducting order parameter  originate with strong frustration due to interband Josephson-couplings. A novel chiral metallic phase found in previous works using large scale Monte-Carlo computations, is not obtained either within the single-site mean-field theory or the improved cluster mean-field theory of order parameter fluctuations. In three-dimensional superconductors, this unusual metallic phase originates with gauge-field fluctuations.  
\end{abstract}

\maketitle

\section{Introduction}

Strong fluctuation effects in condensed matter systems typically manifest themselves in low dimensions at any nonzero temperature, where fundamental theorems~\cite{mermin-wagner,hohenberg,coleman} prevent the breaking of continuous symmetries, such as the loss of translational, rotational, as well as  local and global $\groupU$ symmetries. The first is relevant for freezing of liquids into crystals with long-range order, the second is relevant for ordering in magnets, while the last two give rise to superconductivity and superfluidity. In three dimensions, mean-field theories, where fluctuation effects are often ignored, have met with much success, notably in low-temperature superconductors arising out of good metals.~\cite{schrieffer} This is true even when one attempts to describe the phase transition from the superconducting to the normal metallic state. 

The dominant fluctuations in a strong type-II superconductor/superfluid are phase fluctuations of the order parameter.~\cite{kivelson,tesanovic,nguyen,qed3} The phase-stiffness is governed by the inverse square of the magnetic penetration length $\lambda$, which is small for superconductors with a large Ginzburg-Landau parameter $\kappa = \lambda/\xi$, such as the high-$T_\text{c}$ cuprates or the superconducting pnictides. Here $\xi$ denotes the coherence length. However, in such systems, these fluctuations typically come into play when studying the phase transitions between the various stable states of the systems, while a mean-field calculation works well in the sense of correctly identifying which possible stable phases the systems can feature. In extreme type-II superconductors, with a large Ginzburg-Landau parameter, fluctuations of the electromagnetic vector potential (gauge-field) may also largely be ignored.

In this paper, we show that in multiband superconductors with three or more superconducting bands crossing the Fermi surface, fluctuation effects may be so strong that a simple mean-field calculation fails not only in describing the phase-transitions between the various stable states of the system, but also fails in correctly identifying which possible stable states the system can have. We do this by carrying out single-site and cluster mean-field calculations,\cite{Oguchi,Yamamoto} and compare them to results of large-scale Monte-Carlo simulations, going well beyond what has previously been obtained in the literature.~\cite{PhysRevB.88.094412,PhysRevB.89.104509} In so doing, we identify a source of strong fluctuation effects other than low dimensionality, namely frustration in the phases of the superconducting order parameters due to interband Josephson couplings.~\cite{nagaosa,weston,PhysRevB.89.104509} Examples of such systems are heavy fermion and iron pnictide superconductors.~\cite{seyfarth,hosono}


\section{Reduced Three-Band Model and Its Interpretation}
A standard Ginzburg-Landau theory of an $n$-band superconductor with intra- and intercomponent density-density interactions, and 
inter-component Josephson interactions, is defined by the energy density function (in natural units where we set $\hbar=c=1$)  
\begin{eqnarray}
\mathcal{H} & = & \sum_{\alpha}\frac{1}{2m_\alpha} \left|  \left( \frac{\nabla_\mu}{\i} - e \v{A} \right) \psi_{\alpha} \right|^2  
+ \frac{1}{2} (\nabla \times \v{A})^2 \nonumber \\ 
 & + &\sum_{\alpha', \alpha} \left( u_{\alpha \alpha'} |\psi_\alpha|^2 |\psi_{\alpha'}|^2 + r_{\alpha \alpha'} \psi_{\alpha} \psi^{*}_{\alpha'} \right).
\label{eq:GLTheory}
\end{eqnarray}   
Here, $\alpha,\alpha' \in (1,..n)$ are band-indices, $m_\alpha$ is the mass of the Cooper-pairs originating in band $\alpha$, 
$r_{\alpha,\alpha'}$ represents a term governing the density of Cooper-pairs when 
$\alpha = \alpha'$, $r_{\alpha,\alpha'}$ represents an intercomponent Josephson-coupling when $\alpha \neq \alpha'$, 
and $u_{\alpha \alpha'}$ is the strength of the density-density interactions. 
Furthermore, $\v{A}$ is a fluctuating gauge-field, and $e$ is the charge, here taken to be the same for all components. 
$\psi_\alpha = |\psi_\alpha| \exp(i \theta_\alpha)$ is the complex order-parameter of component $\alpha$, and $\theta_\alpha$ 
is its associated phase. In the moderate to strong type-II regime, where amplitude fluctuations of the order parameter may be 
neglected, the model simplifies to
\begin{multline}
 \mathcal{H}  = \sum_{\alpha} \frac{|\psi_\alpha|^2}{2 m_\alpha} \left(\nabla_\mu \theta_\alpha - e \v{A} \right)^2  +  \frac{1}{2} (\nabla \times \v A)^2 \\
  + \sum_{\alpha' \neq \alpha} r_{\alpha \alpha'} |\psi_{\alpha}| |\psi_{\alpha'}| \cos(\theta_\alpha - \theta_{\alpha'}).  
\label{eq:LondonTheory}
\end{multline}

The lattice version of an $n$-band superconductor in the London limit is given by (when the energy density is summed over 
the entire lattice)~\cite{PhysRevB.88.094412,PhysRevB.89.104509}
\begin{multline}
H = - \sum_{i,\mu,\alpha}  a_\alpha \cos\left(\Delta_\mu \theta_{\alpha,i} - A_{\mu,i}\right) \\
+ \sum_{i,\alpha' > \alpha} g_{\alpha \alpha'} \cos\left(\theta_{\alpha,i} - \theta_{\alpha',i} \right) \\
+ q \sum_{i,\lambda} \Bigl(\sum_{\mu,\nu} \epsilon_{\lambda \mu \nu}\Delta_{\mu}A_{\nu,i}\Bigr)^2.
\label{eq:H_gauge}
\end{multline}
Here, $i \in \set{1,2,\ldots,N=L^3}$ denotes sites of position $\v r_i$ on a lattice of size $L \times L \times L$. $\Delta_{\mu}$ is a difference operator (discrete \enquote{differentiation}; we set the lattice constant to unity) in spatial direction $\mu \in \set{1,2,3}$: $\Delta_{\mu} \theta_{i} \equiv \theta_{\v r_i+\v{e}_\mu} - \theta_{\v r_i}$ (assuming periodic boundary conditions). We may, without loss of generality, choose $a_{1} = 1$, and $a_{\alpha} \in (0,1]$ for $\alpha > 1$. 
Moreover $g_{\alpha \alpha'} \equiv r_{\alpha \alpha'} |\psi_{\alpha}| |\psi_{\alpha'}|$ are renormalized interband Josephson couplings. We have rescaled the gauge field $(A_1,A_2,A_3) = \v A \gets e\v A$ and introduced $q \equiv 1/(2e^2)$. In these units, $q$ parametrizes the London penetration depth of the superconductor. $\epsilon$ is the totally antisymmetric Levi-Civita tensor, with $\lambda, \mu, \nu \in \set{1,2,3}$ as indices. 

When the Josephson couplings $g_{\alpha\alpha'}$ are all positive, each Josephson term by itself prefers to lock phase differences to $\cpi$. For three phases or more, the system is generically frustrated.~\cite{nagaosa,maiti,cp22,PhysRevB.89.104509} {In the ground state it may select one of two possible, inequivalent phase lockings, as illustrated in \cref{fig:gr_state_Z2} for the three band case. By choosing one of these phase locking patterns the system breaks time reversal ($\groupZ$) symmetry.~\cite{nagaosa,zlatko,maiti,johan3,PhysRevB.89.104509}

For the parameters where the model breaks $\groupUZ$ symmetry, it allows topological excitations in the form of domain walls in the $\groupZ$ sector, as well as composite vortices in the $\groupU$ sector.~\cite{cp22,PhysRevB.89.104509} In the composite vortices, all the phases wind by $2\cpi$ and thus they do not carry a topological charge in the $\groupZ$ sector. Thus, proliferation of such vortices cannot disorder phase difference and therefore the system can in principle have a state with broken $\groupZ$ symmetry, but with restored $\groupU$ symmetry. Since in this model there is also a nontrivial interaction between the topological defects in the $\groupU$-sector, i.e. the vortices, and the topological defects in the $\groupZ$-sector, i.e. the domain walls, it requires careful numerical examination under what conditions such a phase may occur (for detailed discussion of vortex and domain wall solutions and their interaction see Ref. \onlinecite{cp22}).

\begin{figure}[ht]
  \subfloat[Phases of the field.\label{fig:phase_definition}]{
    \def\svgwidth{0.4\columnwidth}
  \begingroup%
  \makeatletter%
  \providecommand\color[2][]{%
    \errmessage{(Inkscape) Color is used for the text in Inkscape, but the package 'color.sty' is not loaded}%
    \renewcommand\color[2][]{}%
  }%
  \providecommand\transparent[1]{%
    \errmessage{(Inkscape) Transparency is used (non-zero) for the text in Inkscape, but the package 'transparent.sty' is not loaded}%
    \renewcommand\transparent[1]{}%
  }%
  \providecommand\rotatebox[2]{#2}%
  \ifx\svgwidth\undefined%
    \setlength{\unitlength}{135.16518555bp}%
    \ifx\svgscale\undefined%
      \relax%
    \else%
      \setlength{\unitlength}{\unitlength * \real{\svgscale}}%
    \fi%
  \else%
    \setlength{\unitlength}{\svgwidth}%
  \fi%
  \global\let\svgwidth\undefined%
  \global\let\svgscale\undefined%
  \makeatother%
  \begin{picture}(1,0.59557126)%
    \put(0,0){\includegraphics[width=\unitlength]{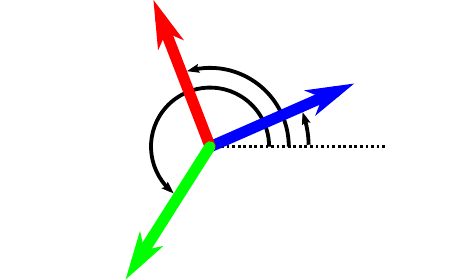}}%
    \put(0.68813623,0.30221329){\color[rgb]{0,0,0}\makebox(0,0)[lb]{\smash{$\theta_{1}$}}}%
    \put(0.5098767,0.4595013){\color[rgb]{0,0,0}\makebox(0,0)[lb]{\smash{$\theta_{2}$}}}%
    \put(0.31188689,0.31269923){\color[rgb]{0,0,0}\makebox(0,0)[rb]{\smash{$\theta_{3}$}}}%
  \end{picture}%
  \endgroup%
  }\\
  \subfloat[$+1$\label{fig:3C_ground_state_1}]{\includegraphics[width=0.4\columnwidth]{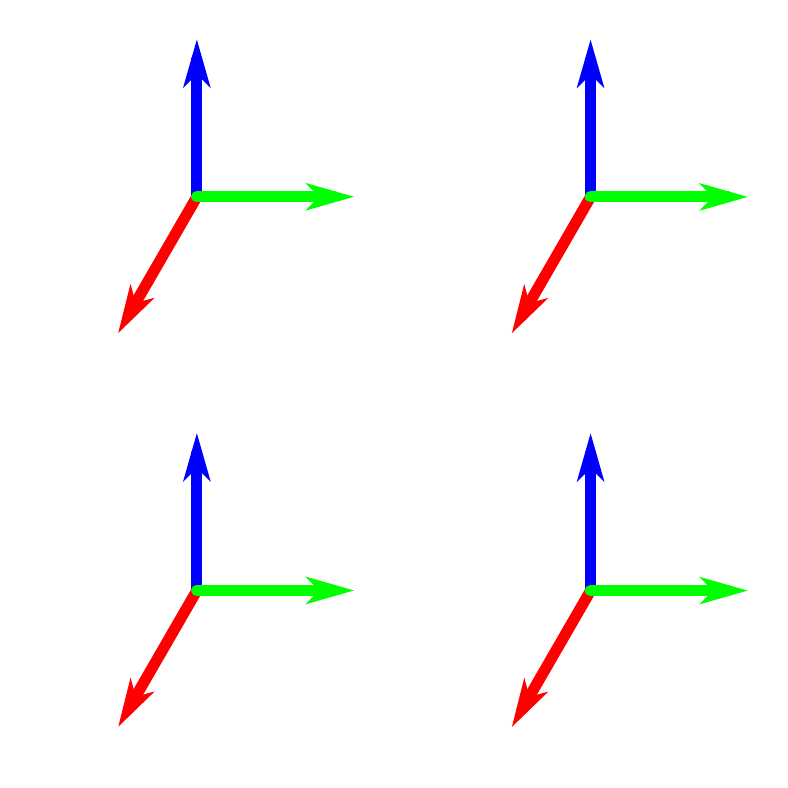}}
  \qquad
  \subfloat[$-1$\label{fig:3C_ground_state_2}]{\includegraphics[width=0.4\columnwidth]{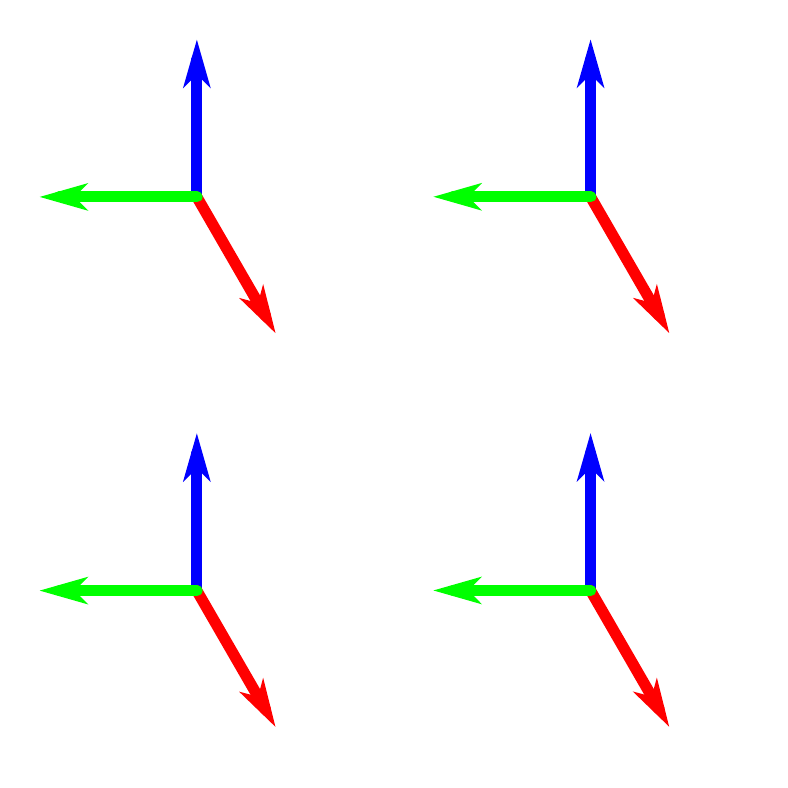}}
  \caption{(Colors online) Illustration of a $n=3$ band superconductor. The arrows in panel \protect\subref{fig:phase_definition} $(\textcolor{blue}{\longrightarrow},\textcolor{red}{\longrightarrow},\textcolor{green}{\longrightarrow})$ correspond to $(\theta_1,\theta_2,\theta_3)$. Panels \protect\subref{fig:3C_ground_state_1} and \protect\subref{fig:3C_ground_state_2} show examples of phase configurations for the two $\groupZ$ symmetry classes of the ground states, shown on a $2\times 2$ lattice 
  of a planar slice of the system. Here $g_{12} > g_{23} > g_{13} > 0$. The spatial contribution to the energy is minimized by making the spatial gradient zero (hence breaking the global $\groupU$ symmetry). Then there are two classes of phase configurations, one with chirality +1 and one with chirality -1, minimizing the energy associated with the interband interaction. The chirality is defined as $+1$ if the phases (modulo $2\cpi$) are cyclically ordered $\theta_1<\theta_2<\theta_3$, and $-1$ if not.}  
\label{fig:gr_state_Z2}
\end{figure}

In the limit $e \to 0 \Leftrightarrow q \to \infty$, where fluctuations in the gauge field may be neglected, the model is reduced to
\begin{multline}
H = - \sum_{i,\mu,\alpha} a_\alpha \cos\left(\Delta_\mu \theta_{\alpha,i} \right) \\
+ \sum_{i,\alpha' > \alpha} g_{\alpha \alpha'} \cos\left(\theta_{\alpha,i} - \theta_{\alpha',i} \right).
\label{eq:H_basic}
\end{multline}

We next proceed to simplifying \cref{eq:H_gauge} further, in a way that is appropriate for these types of systems. By letting $g_{\alpha \alpha'} \to \infty$ {in the lattice London model} such that the ratio $g_{\alpha \alpha'}/g_{\beta\beta'}$ is finite ($\alpha,\alpha',\beta,\beta'$ being band indices), we may derive a \enquote{reduced} version of the model given by \cref{eq:H_gauge,eq:H_basic}, for which the intercomponent phase fluctuations are essentially suppressed. Namely, the \enquote{phase star} of a lattice site locks into one of the two possible $\groupZ$ configurations minimizing the contribution from the Josephson term in the Hamiltonian. That is, in this approximation the phase differences can have only two values. The $\groupZ$ domain wall then represents a change of the phase difference over one lattice spacing.

For the case without a fluctuating gauge-field, the reduced lattice London model is given by a rather unusual coupled Ising-XY type of model
\begin{multline}
 H = -\sum_{i,\mu}\big[ (1 + K_1\sigma_i \sigma_{i+\mu})\cos(\Delta_\mu \theta_{i}) \\
 + K_2 (\Delta_{\mu}\sigma_i)\sin(\Delta_\mu \theta_{i}) \big].
 \label{eq:H_reduced}
\end{multline}
For details of the derivation of the somewhat unfamiliar model \cref{eq:H_reduced} from the more familiar model \cref{eq:H_basic}, see Appendix A of Ref.~\onlinecite{PhysRevB.89.104509}. In \cref{eq:H_reduced} $\sigma_i \in \set{-1,+1}$ denotes the $\groupZ$ chirality of the \enquote{phase star}, $\theta_i \equiv \theta_{1,i}$ its overall orientation, and
\begin{align}
 K_1 \equiv& \frac{\sum_{\alpha>1} a_\alpha\bigl[1-\cos(2\phi_{\alpha})\bigr]}{2 + \sum_{\alpha>1} a_\alpha \bigl[1+\cos(2\phi_{\alpha})\bigr]} \label{eq:K1} \\
 K_2 \equiv& \frac{\sum_{\alpha>1} a_\alpha \sin(2\phi_{\alpha})}{2 + \sum_{\alpha>1} a_\alpha \bigl[1+\cos(2\phi_{\alpha})\bigr]}.
 \label{eq:K2}
\end{align}
$\phi_{\alpha}$ is the -- now fixed -- phase difference between component 1 and component $\alpha$: $\phi_\alpha \equiv \theta_{\alpha,i}-\theta_{1,i}$. The $\phi_{\alpha}$'s are determined by the ratios $g_{\alpha \alpha'}/g_{\beta\beta'}$ of the Josephson-couplings. For site-independent Josephson-couplings, the $\phi_\alpha$'s are also site-independent. Inspecting \cref{eq:K1,eq:K2}, we see that $K_1$ and $K_2$ are measures of how the phase differences are distributed in the phase stars: In the three component case, with $a_{\alpha} = 1$, $K_1 = 0$ denotes the case $\phi_2 = \phi_3 = \cpi$, i.e. where the repulsion between component 1 and 2 and 3 dominate over the repulsion between components 2 and 3. $K_1 = 1$ if the phases are maximally symmetrically distributed, $\phi_2 = -\phi_3 = 2\cpi/3$. $K_1 = 2$ when the repulsion between 2 and 3 dominates, i.e. $\phi_{2} = -\phi_3 = \cpi/2$. Similarly, $K_2$ is a measure of the \enquote{skewness} of the phase star, with $K_2 = 0$ when $\phi_2 = -\phi_3$ in the cases above, and $K_2 \neq 0$ if $\phi_2 \neq -\phi_3$. 

The term $(1 + K_1\sigma_i \sigma_{i+\mu})\cos(\Delta_\mu \theta_{i})$ promotes a fully uniform superconducting phase where all phases of the three 
components of the superconducting order parameter are phase-locked and $\groupU$-ordered. The parameter $K_1$ plays the role of suppressing the formation of superconducting domains of opposite chirality. The term $K_2 (\Delta_{\mu}\sigma_i) \sin(\Delta_\mu \theta_{i})$, on the other hand, tends to promote a phase which is non-uniform both in the $\groupZ$- and $\groupU$-sectors. That is, the parameter $K_1$ tends to suppress phase fluctuations of the overall phase-locked star, while the parameter $K_2$ tends to enhance phase-fluctuations of the phase-locked star as well as introducing domains of superconducting order with opposite chirality. Effectively therefore, the first term in \cref{eq:H_reduced} 
suppresses phase-fluctuations, while the second term enhances phase-fluctuations and reduces the energy of $\groupZ$-domain walls in the system. 

If $K_1$ and $K_2$ are treated as free parameters, the model \cref{eq:H_reduced} in principle allows a uniform as well as staggered ordering of the $\groupZ$ $\sigma_i$-variables on the lattice, in addition to the disordered state. A uniform ordering means that the phases illustrated in \cref{fig:phase_definition} have the same chirality throughout the lattice, while a staggered ordering means that the chirality alternates on some length scale of the lattice. We will refer to the former as \enquote{ferromagnetic} ordering in the $\groupZ$ sector, while the latter will be referred to as \enquote{antiferromagnetic}. We should bear in mind, however, that for an $n$-band London superconductor with inter-band Josephson-coupling, there is a constraint on the parameters $(K_1,K_2)$ which prevents the \enquote{antiferromagnetic} from taking place. See Ref.~\onlinecite{PhysRevB.89.104509} for details on the derivation of \cref{eq:H_reduced} and the physical domain of the $K_1,K_2$ plane. 

One may ask if the results obtained using \cref{eq:H_reduced}, to be presented in \cref{fig:phase diagrams} c) and d) below, are an artifact of the rigid phase-star approximation encoded in \cref{eq:H_reduced}, and whether essentially the same results would be obtained were the model \cref{eq:H_gauge} to be used. In a previous work,~\cite{PhysRevB.89.104509} we have compared results obtained using \cref{eq:H_gauge,eq:H_reduced} for $K_2=0$. (In the present work, we study the model also for finite $K_2$). The results based on using \cref{eq:H_reduced} are qualitatively and quantitatively very similar to those based on \cref{eq:H_gauge}. We thus believe that that the results based on \cref{eq:H_reduced} are faithful representations of those that would be obtained using \cref{eq:H_gauge}. This is also what one would conclude
on general grounds based on an analysis of the scaling dimension of the Josephson-coupling. 

Up to an overall scaling factor, \cref{eq:H_reduced} may also be written on a somewhat more familiar form of a coupled Ising-XY model,~\cite{PhysRevB.90.134512}
\begin{equation}
 H = - \sum_{i,\mu} (1 + J\sigma_i\sigma_{i+\mu}) \cos\bigl(\Delta_{\mu}\theta_i - \gamma(\sigma_i,\sigma_{i+\mu})\bigr)
 \label{eq:H_reduced_alternative}
\end{equation}
where $J = \frac{W}{1 + \sqrt{1 - W^2}}$, $W \equiv \frac{2(K_1 - {K_2}^2)}{1 + {K_1}^2 + 2{K_2}^2}$, and
\begin{equation}
 \gamma(\sigma_i,\sigma_j) \equiv \begin{cases}
      0 & \sigma_i = \sigma_j \\
      \pm \arctan\left[\frac{2K_2}{1 - K_1 }\right] &\sigma_i = -\sigma_j = \pm 1
     \end{cases}
\label{eq:XY-Ising}		
\end{equation}

We emphasize that, although the model given in \cref{eq:H_reduced,eq:H_reduced_alternative} may look unfamiliar in the context of multi-band superconductivity, they are straightforwardly derived from a familiar Ginzburg-Landau theory for a three-band superconductor with interband Josephson-couplings in the London-approximation, \cref{eq:H_gauge}, in the limit of strong Josephson-couplings. The emergence of the Ising-variables $\sigma_i$ associated with two distinct chiralities of the 
three-phase-star in Fig. \ref{fig:gr_state_Z2}, is the positive sign of the interband Josephson-couplings in Eq. \ref{eq:GLTheory}. The effective stiffness of 
the domain-walls in Eq. \ref{eq:H_reduced_alternative} is determined by the parameter $J= J(K_1,K_2)$. The fluctuating $\groupZ$ \enquote{gauge-field} $\gamma(\sigma_i,\sigma_{i+\mu})$ appearing in \cref{eq:H_reduced_alternative} and defined in \cref{eq:XY-Ising} is another manifestation of interaction between the superconducting domains and the fluctuating domain walls separating domains of opposite chirality. Namely, any change in chirality by necessity leads to a local fluctuation in phase-gradients. This has a similar effect as a gauge-field on the supercurrents $\Delta_{\mu}\theta_i$. The coefficient $1 + J\sigma_i\sigma_{i+\mu}$ in Eq. \ref{eq:H_reduced_alternative}  acts as an effective
bare superfluid density, while the \enquote{gauge-field}-fluctuations lead to a reduction of this stiffness. Eq. \ref{eq:H_reduced_alternative} thus effectively 
describes a one-component extreme type-II superconductor associated with the overall fluctuations of the three-phase-star, in the presence of an emergent fluctuating 
$\groupZ$ \enquote{gauge-field} associated with fluctuating domain wall separating domains of opposite chirality. 
A reduced model including a $\groupU$ gauge-field is obtained by replacing $\Delta_{\mu}\theta_i$ by $\Delta_{\mu}\theta_i - A_{\mu,i}$ in \cref{eq:H_reduced} or \cref{eq:H_reduced_alternative} and adding a Maxwell term $q \sum_{i,\lambda} \bigl(\sum_{\mu,\nu} \epsilon_{\lambda \mu \nu}\Delta_{\mu}A_{\nu,i}\bigr)^2$ to the Hamiltonian. This would be appropriate for moderate type-II three-band superconductors. 

\section{Results}

The free energy density of the reduced model in the mean-field approximation is given by (see \cref{app:MF}) 
\begin{multline}
  f = M \bigg[r \frac{I_1(\beta M r)}{I_0(\beta M r)} - \frac{1}{2}\left(\frac{I_1(\beta M r)}{I_0(\beta M r)}\right)^2 \bigg] \\
  + \beta\inv\left[s_{\groupZ}(m) - \ln \bigl( I_0(\beta M r) \bigr) \right],
 \label{eq:f4}
\end{multline}
with $ M = 1 + K_1 m^2, \quad m \equiv \tfrac{1}{2}(m_\text{A} + m_\text{B})$ when the $\groupZ$ sector will order \enquote{ferromagnetically}, and
$M = \sqrt{(1 - K_1 m^2)^2 + 4{K_2}^2 m^2}, \quad m \equiv \tfrac{1}{2}(m_\text{A} - m_\text{B})$, 
when the ordering is \enquote{antiferromagnetic}. Antiferromagnetic $\groupZ$ ordering can take place when $K_1 < {K_2}^2$, i.e. for large enough $K_2$. This situation is unphysical when viewing the reduced model as a limiting case of the multiband London model,~\cite{PhysRevB.89.104509} i.e. when $K_1$ and $K_2$ are determined by \cref{eq:K1,eq:K2}, but is included here for the sake of completeness. $m_\text{A}$ and $m_\text{B}$ are the Ising-type magnetizations on sublattices A and B of the bipartite lattice, while $r$ is the condensate density ($\groupU$ order parameter). Furthermore, the $I_l$'s are modified Bessel functions of order $l$ and $s_{\groupZ}(m) \equiv \left( \frac{1 + m}{2} \right)\ln\left( \frac{1 + m}{2} \right) + \left( \frac{1 - m}{2} \right)\ln\left( \frac{1 - m}{2} \right)$. An immediate consequence of this mean-field form is that when $r=0$, we have $f = \beta^{-1} s_{\groupZ}(m)$, which has a global minimum at $m=0$. Thus, at the mean-field level, there can be no broken $\groupZ$ symmetry in a $\groupU$-symmetric (metallic) state. As we shall see, strong fluctuation effects alter this picture quite drastically, \emph{even in three dimensions}.

With $m = 0$, $M=1$ and the free energy \cref{eq:f4} reduces (up to a constant term) to that of the XY model, 
\begin{equation}
 f_\text{XY} = r \frac{I_1(\beta r)}{I_0(\beta r)} - \frac{1}{2}\left(\frac{I_1(\beta r)}{I_0(\beta r)}\right)^2 - \beta\inv \ln \bigl( I_0(\beta r)\bigr),
 \label{eq:f5}
\end{equation} 
which displays a second order phase transition at $\beta_\text{c} = 2$.

\begin{figure*}[]
\subfloat[The mean-field phase diagram. \label{fig:MF_phasediagram}]{\includegraphics{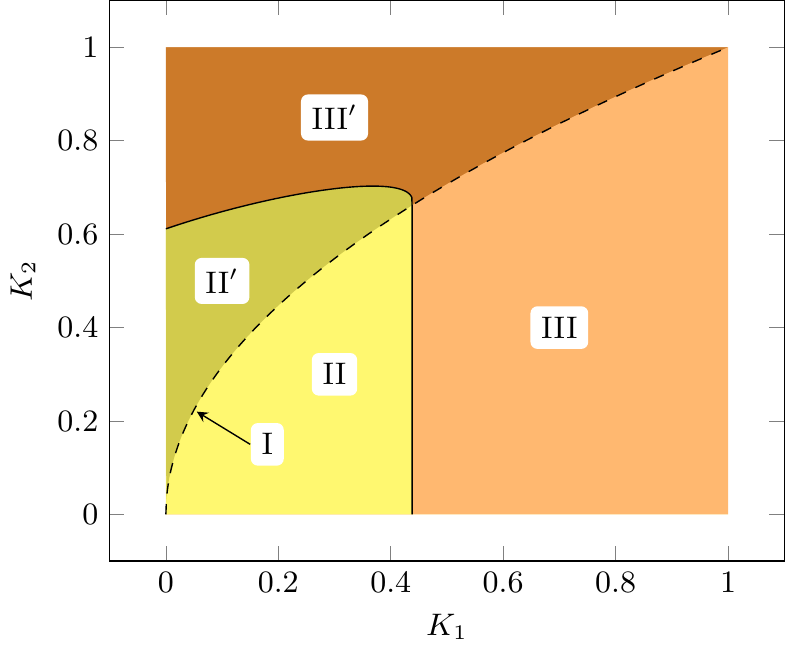}}
\subfloat[The cluster mean field phase diagram. \label{fig:CMF_phasediagram}]{\includegraphics{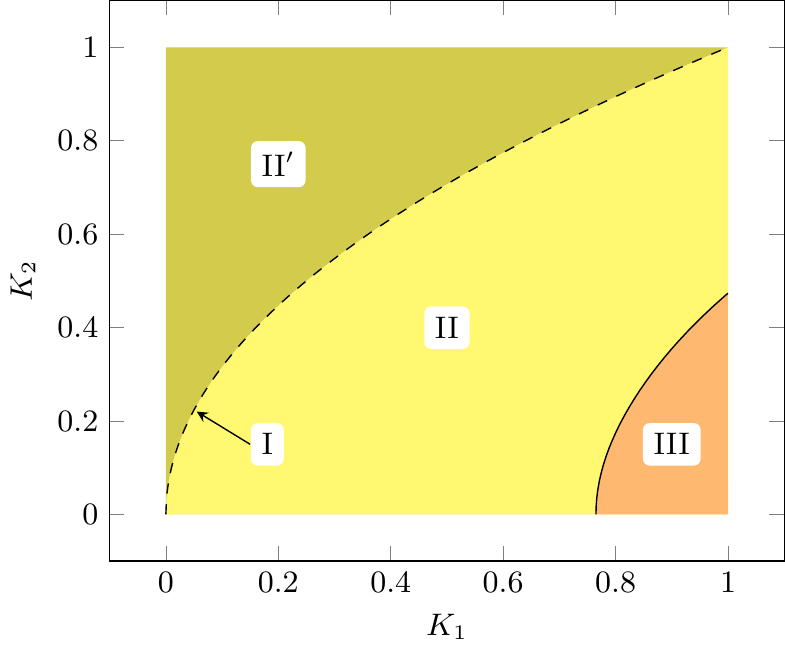}} \\
\subfloat[The phase diagram without a fluctuating gauge field. \label{fig:fluctuating_nogauge}]{\includegraphics{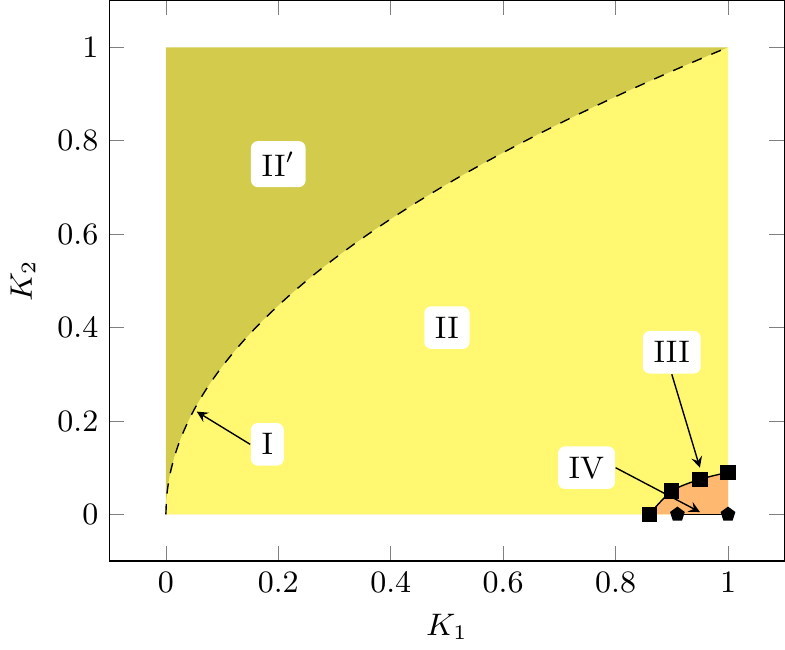}}
\subfloat[The phase diagram with a fluctuating gauge field. \label{fig:fluctuating_gauge}]{\includegraphics{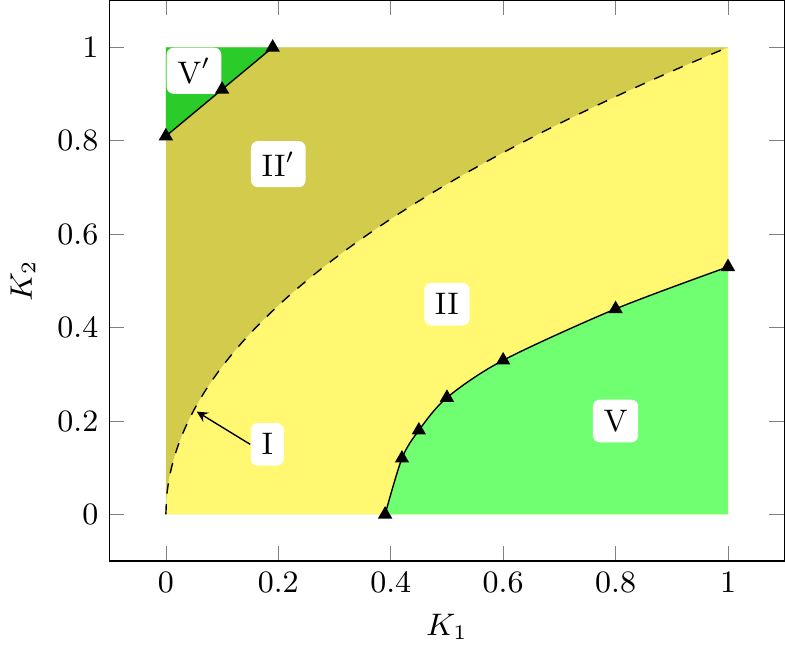}}
\caption[]{
\protect\subref{fig:MF_phasediagram} The mean field phase diagram of the $K_1 K_2$ model, based on minimizing the free energy, \cref{eq:f4}. \protect\subref{fig:CMF_phasediagram} The cluster mean field phase diagram of the $K_1K_2$ model based on a $2\times 2 \times 2$ cluster. \protect\subref{fig:fluctuating_nogauge} The phase diagram of the $K_1 K_2$ model without a fluctuating gauge field, \cref{eq:H_reduced}. The plot is based on Monte Carlo simulations with $L=40$, except for $K_2=0$, where $L=50$ was used. The markers (\begin{tikzpicture}
\draw[gray] (0,0) node {};
\pgfplothandlermark{\pgfuseplotmark{square*}}
\pgfplotstreamstart
\pgfplotstreampoint{\pgfpointorigin}
\pgfplotstreamend
\pgfusepath{stroke}
\end{tikzpicture},\begin{tikzpicture}
\draw[gray] (0,0) node {};
\pgfplothandlermark{\pgfuseplotmark{pentagon*}}
\pgfplotstreamstart
\pgfplotstreampoint{\pgfpointorigin}
\pgfplotstreamend
\pgfusepath{stroke}
\end{tikzpicture}) indicate points determined through the simulations, while the lines are guides to the eye. \protect\subref{fig:fluctuating_gauge} The phase diagram of the $K_1 K_2$ model with a fluctuating gauge field, $q=0.1$. The plot is based on Monte Carlo simulations with $L=40$. The markers (\begin{tikzpicture}
\draw[gray] (0,0) node {};
\pgfplothandlermark{\pgfuseplotmark{triangle*}}
\pgfplotstreamstart
\pgfplotstreampoint{\pgfpointorigin}
\pgfplotstreamend
\pgfusepath{stroke}
\end{tikzpicture}) indicate points determined through the simulations, while the lines are guides to the eye. Unprimed labels denote a state where the ordering in the $\groupZ$-sector is \enquote{ferromagnetic}, while primed labels indicate that this ordering is \enquote{antiferromagnetic}. The latter is unphysical when viewing the $K_1K_2$ model as a limiting case of the London model. I: The borderline between the ferromagnetic and antiferromagnetic regions. On this borderline, there is only $\groupU$ ordering, no $\groupZ$ ordering. $\text{II}, \text{II}'$: The $\groupZ$ transition is second order and happens when the $\groupU$ sector has already ordered. $\text{III}, \text{III}'$: The $\groupZ$ transition is first order and happens when the $\groupU$ sector has already ordered. The solid line separating the regions $(\text{II}, \text{II}')$ from the regions $(\text{III}, \text{III}')$ indicates the boundary where the phase transition in the $\groupZ$ sector changes from second order to first order. For I-III the $\groupU$ transition is second order. IV: The $\groupZ$ and $\groupU$ sectors order at the same time through a first order transition. V: The $\groupZ$ sector orders before the $\groupU$ sector, i.e. the system displays a region of an anomalous, $\groupZ$ broken metallic state. Both sectors order through second order transitions. Note that in the context of a general coupled $XY$-Ising model, the region above the dotted line shows antiferromagnetic ordering, while in the context of a three-band London superconductor with interband Josephson coupling, this region of the phase diagram is unphysical. See text for details.
}
\label{fig:phase diagrams}
\end{figure*}

In \cref{fig:MF_phasediagram}, we show the phase diagram of the model \cref{eq:H_reduced}, based on the mean-field free energy \cref{eq:f4}. The dashed line is the separatrix between \enquote{ferromagnetic} and \enquote{antiferromagentic} $\groupZ$-ordering in the Ising-pseudospin-sector. Precisely on the dotted line, the system never orders in the $\groupZ$ sector, since the energy of the $\groupZ$ domain walls vanishes there. The solid black line is the separatrix in $(K_1,K_2)$-space between a second-order and first-order phase-transition in the $\groupZ$-sector, i.e. a tricritical boundary line. 

In \cref{fig:CMF_phasediagram}, we show an improved cluster-mean field phase diagram (see \cref{app:CMF}) based on a $2 \times 2 \times 2$ cluster where fluctuations are allowed. This represents a first step towards including fluctuation corrections to the mean-field phase-diagram of \cref{fig:MF_phasediagram}, which is essentially based on a $1 \times 1 \times 1$ cluster. We see that the tricritial boundary line separating $(\text{II},\text{II}')$ from $(\text{III},\text{III}')$ is pushed considerably further away from the origin of the $K_1,K_2$ plane, due to fluctuation effects even at this level. This result in itself indicates that fluctuation effects are strong in these systems.  

\Cref{fig:fluctuating_nogauge} shows the phase diagram for the case with no fluctuating gauge field, obtained by Monte Carlo simulations (see \cref{app:MC}). The tricritical boundary line is altered considerably compared to what is found in \cref{fig:MF_phasediagram} and \cref{fig:CMF_phasediagram}. Furthermore, for $K_2=0$ and sufficiently large $K_1$ values, the transitions in the $\groupZ$ and $\groupU$ sector merge into a single joint first order transition not seen in the mean field case.
Note that the results shown in \cref{fig:CMF_phasediagram} qualitatively compare well with the results shown \cref{fig:fluctuating_nogauge}. Although the differences between these results and those shown in \cref{fig:MF_phasediagram} are large, it is encouraging that the refined $2 \times 2 \times 2$ cluster mean-field analysis already seems converged reasonably well to the numerical results.    

Adding a fluctuating gauge field, the discrepancy between the true (Monte Carlo) and the mean-field picture is even more striking. \Cref{fig:fluctuating_gauge} gives the phase diagram when $q=0.1$. The $\groupZ$ transition now remains second order in the entire phase diagram. A new, $\groupU$-symmetric (metallic), but $\groupZ$ broken (chiral) state emerges. Typically, one expects mean-field calculations in a three-dimensional system to at least yield a correct phase diagram. Here, we see that strong intrinsic fluctuation effects in multi-band superconductors with more than two bands alter this basic picture, and that some level of fluctuations must be taken into account to obtain a reasonably correct phasediagram. 

\section{Summary and conclusions}

Previous works have found a chiral metallic state of Josephson-coupled three-band superconductors, such as the iron-pnictides, in large-scale Monte Carlo simulations.~\cite{PhysRevB.88.094412,PhysRevB.89.104509} In this paper, we have investigated whether or not mean-field theories are capable of yielding such novel phases
in three-dimensional superconductors, where fluctuation effects normally are considered to be moderate.  

To this end, we have computed the single-site and cluster mean-field phase-diagrams of the model \cref{eq:H_gauge}, in the representation \cref{eq:H_reduced}, and compared with available large-scale Monte Carlo results taking fully into account fluctuations in the problem. The single-site and cluster mean-field calculations we have performed have taken into account {\it order-parameter} fluctuations, but not gauge-field fluctuations. The main finding is that a simple single-site mean-field calculation, \cref{fig:phase diagrams}, does not capture the correct phase diagram of the phase-fluctuating system, even in the extreme type-II limit where there are no gauge-field fluctuations. However, upon introducing a refined analysis involving cluster mean-field calculations, already a $2 \times 2 \times 2$-cluster mean-field calculation improves the results considerably, yielding a phase diagram which is qualitatively correct in the extreme type-II limit, when compared with large-scale Monte-Carlo calculations. It thus appears that including order-parameter fluctuations at this level produces reliable results in the extreme type-II limit. Thus, we have demonstrated that i) fluctuation effects are strong in these compounds, and ii) relatively modest refinements beyond the simple mean-field approaches yield results in surprisingly good agreement with results obtained in large-scale computations. However, gauge-field fluctuations are required in order to produce a chiral metallic phase.~\cite{PhysRevB.88.094412,PhysRevB.89.104509}

While strong order-parameter fluctuation effects are well known in superconductors and superfluids in two dimensions,~\cite{mermin-wagner,hohenberg,coleman} it is much more uncommon to see such strong fluctuation effects in higher-dimensional systems. They originate with strong frustration due to interband Josephson-couplings.   

T.A.B. thanks NTNU for financial support. A.S. was supported by the Research Council of Norway, through Grants 205591/V20 and 216700/F20. AS thanks the Aspen Center for Physics  (NSF Grant No 1066293) for hospitality during the initial stages of this work. This work was also supported through the Norwegian consortium for high-performance
computing (NOTUR).

\appendix

\section{Mean field calculations}\label{app:MF}
Obtaining an expression for the (mean field) free energy of the model as a function of the order parameters of the symmetry sectors, yields the (mean field) phase diagram, \cref{fig:MF_phasediagram}.

\subsection{The free energy}
Our goal is to derive a mean field free energy density for the lattice model given by the Hamiltonian
\begin{equation}
 H = \sum_{\langle i,j\rangle} h_{ij} 
 \label{eq:H_reduced_A}
\end{equation}
where\footnote{Since we are dealing with a mean field model, the gauge field is fixed and may be removed by selecting a proper gauge. Hence, up to an irrelevant constant, the expression is independent of $q$.}
\begin{align}
 h_{ij} &= -\big[ (1 + K_1\sigma_i \sigma_j)\cos(\theta_i - \theta_j) \nonumber\\
 &\hphantom{= -\big[} + K_2(\sigma_i - \sigma_j)\sin(\theta_i - \theta_j) \big]
 \label{eq:h_real} \\
 &= -\tfrac{1}{2} \big[1 + K_1\sigma_i \sigma_j - \i K_2(\sigma_i -\sigma_j)\big]\e{\i\theta_i}\e{-\i\theta_j} + \text{c.c.} \label{eq:h_complex}
\end{align}
Here, we have introduced the notation $\langle i,j \rangle$ for the nearest neighbor sites $i$ and $j$ (as a less cluttered alternative to the pair $(i,i+\mu)$). The lattice is bipartite with coordination number $z$ and volume $N$. We will assume that $K_1>0$.

In general, the free energy of a system may be written as~\cite{Arovas}
\begin{equation}
 F = E - TS = \Tr [\rho H] + T \Tr [\rho \ln \rho],
 \label{eq:F_general}
\end{equation}
where the density matrix $\rho$ is subject to the normalization constraint
\begin{equation}
 \Tr \rho = 1.
 \label{eq:norm}
\end{equation}
The true \emph{equilibrium} free energy is the minimum of \cref{eq:F_general} over all possible $\rho$'s. Here, we restrict ourselves to the (tractable) subset of density matrices being a direct product of independent, single site contributions:
\begin{equation}
 \rho = \bigotimes_{i} \rho_{i},
\end{equation}
In other words: we ignore fluctuation effects.

Due to symmetry, the mean field density matrices of all sites of a sublattice must be identical. The density matrix of the other sublattice may however be different, as we can expect both \emph{canted} ordering in the $\groupU$ sector as well as \emph{antiferromagnetic} ordering in the $\groupZ$ sector. Hence, we write
\begin{equation}
 \rho = \bigotimes_{2i} \rho_{\text{A}}\otimes \rho_{\text{B}},
 \label{eq:subdivision}
\end{equation}
where the two sublattices are labeled A and B.

Using \cref{eq:H_reduced_A,eq:subdivision} we get
\begin{align}
 \Tr[\rho H] &= \frac{N}{2}\frac{z}{2}\Bigl(\Tr[\rho_\text{A}\otimes\rho_\text{B}h_{\text{AB}}] + \Tr[\rho_\text{A}\otimes\rho_\text{B}h_{\text{BA}}]\Bigr) \nonumber\\
 &= \frac{Nz}{2}\Tr[\rho_\text{A}\otimes\rho_\text{B}h_{\text{AB}}]
\end{align}
and
\begin{align}
 \Tr [\rho \ln \rho] &= \frac{N}{2}\Bigl(\Tr[\rho_\text{A}\ln \rho_\text{A}] + \Tr[\rho_\text{B}\ln \rho_\text{B}] \Bigr),
\end{align}
leading to a free energy density of
\begin{multline}
 f \equiv \frac{F}{N} = \tfrac{1}{2}z\Tr[\rho_\text{A}\otimes\rho_\text{B}h_{\text{AB}}] + \\
 \tfrac{1}{2}T\Bigl(\Tr[\rho_\text{A}\ln \rho_\text{A}] + \Tr[\rho_\text{B}\ln \rho_\text{B}] \Bigr).
 \label{eq:f_general}
\end{multline}

$\rho_\text{A(B)}$ may be decomposed into density matrices of the $\groupU$ and the $\groupZ$ sector:
\begin{equation}
 \rho_{\text{A(B)}} = \rho_{\groupU,\text{A(B)}} \otimes \rho_{\groupZ,\text{A(B)}},
\end{equation}
where $\rho_{\groupU} = \rho_{\groupU}(\theta)$ and $\rho_{\groupZ} = \rho_{\groupZ}(\sigma)$. From \cref{eq:norm} we immediately see that we can write
\begin{equation}
 \rho_{\groupZ} = \frac{1+m}{2}\delta_{\sigma,1} + \frac{1-m}{2}\delta_{\sigma,-1}
 \label{eq:rho_Z2}
\end{equation}
where $m$ is a parameter (the $\groupZ$ \enquote{magnetization} of the site) to be determined. $\rho_{\groupU}$ is a bit more subtle and will be established in the following free energy minimization.

First, we want to integrate out the $\groupZ$ degrees of freedom. Inserting \cref{eq:h_complex,eq:rho_Z2} into the first term of \cref{eq:f_general}, using that
\begin{equation} 
 \Tr[\sigma \rho] = \Tr[\sigma \rho_{\groupZ}] = \frac{1+m}{2} - \frac{1-m}{2} = m,
\end{equation}
yields
\begin{widetext}
  \begin{equation} \tfrac{1}{2}z\Tr[\rho_\text{A}\otimes\rho_\text{B} h_{\text{AB}}]
  = - \tfrac{1}{4} z \big[1 + K_1 m_\text{A} m_\text{B} - \i K_2( m_\text{A} -m_\text{B})\big] 
  \Tr\big[\rho_{\groupU,\text{A}}\e{\i\theta_\text{A}}\big] \Tr\big[\rho_{\groupU,\text{B}}\e{-\i\theta_\text{B}}\big]  + \text{c.c.}
  \label{eq:f_E}
  \end{equation}
  In the same way,
  \begin{equation}
  \Tr[\rho_\text{A}\ln \rho_\text{A}] + \Tr[\rho_\text{B}\ln \rho_\text{B}] = s_{\groupZ}(m_\text{A}) + s_{\groupZ}(m_\text{B}) + \Tr[\rho_{\groupU,\text{A}}\ln \rho_{\groupU,\text{A}}] + \Tr[\rho_{\groupU,\text{B}}\ln \rho_{\groupU,\text{B}}],
  \label{eq:f_S}
  \end{equation}
  where
  \begin{equation}
   s_{\groupZ}(m) \equiv \tfrac{1}{2}(1+m)\ln\big[\tfrac{1}{2}(1+m)\big] + \tfrac{1}{2}(1-m)\ln\big[\tfrac{1}{2}(1-m)\big].
  \end{equation}
\end{widetext}
To keep notation simple (while still being unambiguous), we omit the subscript $\groupU$ and just write $\rho$ for $\rho_{\groupU}$ from now on.

We may now proceed to determine $\rho$.  Minimizing the the free energy, \cref{eq:F_general}, subject to the normalization constraint \cref{eq:norm}, is equivalent to minimizing the \enquote{extended} free energy density
\begin{equation}
 \tilde{f} = f - \tfrac{1}{2}T[\lambda_\text{A}(\Tr \rho_{\text{A}} - 1) + \lambda_\text{B}(\Tr \rho_{\text{B}} - 1)]
 \label{eq:extended_f}
\end{equation}
without constraints. Here $\lambda_\text{A}$ and $\lambda_\text{B}$ are (conveniently scaled) Lagrange multipliers.

The minimum is found when
\begin{align}
 \pderiv{\tilde{f}}{\lambda_{\text{A(B)}}} &= 0 \label{eq:stationary_lambda}\\
 \fderiv{\tilde{f}}{\rho_{\text{A(B)}}} &= 0 \label{eq:stationary_rho}
\end{align}

Note that for an arbitrary \emph{function} $g$ we have that
\begin{equation}
 \fderiv{}{\rho} \Tr[\rho g] = \fderiv{}{\rho} \int_{-\cpi}^{\cpi} \frac{\diff\theta}{2\cpi} \rho(\theta)g(\theta) = \frac{g}{2\cpi},
\end{equation}
so \cref{eq:stationary_rho} gives
\begin{widetext}
  \begin{equation}
  0 = - \tfrac{1}{2} z \big[1 + K_1 m_\text{A} m_\text{B} - \i K_2( m_\text{A} -m_\text{B})\big] 
  \e{\i\theta_\text{A}} \Tr\big[\rho_{\text{B}}\e{-\i\theta_\text{B}}\big]  + \text{c.c.} + T(\ln \rho_\text{A} + 1 - \lambda_{\text{A}})
  \label{eq:stationary}
  \end{equation}
\end{widetext}
and equivalently for $\text{A} \leftrightarrow \text{B}$. 

Furthermore, there exist two quantities $r \in [0,1]$ and a $\theta_0 \in [-\cpi,\cpi)$ such that
\begin{equation}
 \Tr[\rho \e{\i\theta}] = \int_{-\cpi}^{\cpi} \frac{\diff\theta}{2\cpi} \rho(\theta)\e{\i\theta} = r\e{\i\theta_0}
 \label{eq:trace_exp}
\end{equation}
Since the system is $\groupU$ symmetric, we may choose a coordinate system such that $\theta_{0,\text{A}} = -\theta_{0,\text{B}} = \alpha$. Using this and inserting \cref{eq:trace_exp} into \cref{eq:stationary}, solving for $\rho$, leaves us with
\begin{align}
 \rho_{\text{A}} &= \e{\lambda_{\text{A}}-1}\exp\Big\{\beta r_{\text{B}}[(1 + K_1 m_{\text{A}} m_{\text{B}})\cos(\theta_{\text{A}} + \alpha) \nonumber\\
 &\hphantom{{}=\e{\lambda_{\text{A}}-1}\exp\Big\{}+ K_2(m_{\text{A}} - m_{\text{B}})\sin(\theta_{\text{A}} + \alpha)] \Big\} \\
 \rho_{\text{B}} &= \e{\lambda_{\text{B}}-1}\exp\Big\{\beta r_{\text{A}}[(1 + K_1 m_{\text{A}} m_{\text{B}})\cos(\theta_{\text{B}} - \alpha) \nonumber\\
 &\hphantom{{}=\e{\lambda_{\text{B}}-1}\exp\Big\{}+ K_2(m_{\text{A}} - m_{\text{B}})\sin(\theta_{\text{B}} - \alpha)] \Big\}
\end{align}
where
\begin{equation}
 \beta \equiv \frac{z}{T}.
\end{equation}

The $\lambda$'s are determined by \cref{eq:stationary_lambda}, which is just the normalization constraint, $\Tr \rho = 1$. By integration:
\begin{equation}
 \e{\lambda_{\text{A(B)}}-1} I_0(\beta M r_{\text{B(A)}}) = 1
\end{equation}
where
\begin{equation}
 M \equiv \sqrt{(1 + K_1 m_{\text{A}} m_{\text{B}})^2 + {K_2}^2(m_{\text{A}} - m_{\text{B}})^2},
\end{equation}
and $I_l$ is the $l$'th order modified Bessel function. The final expressions for the $\rho$'s are therefore
\begin{align}
 \rho_{\text{A}} &= I_0(\beta M r_{\text{B}})\inv \exp\Big\{\beta r_{\text{B}}[(1 + K_1 m_{\text{A}} m_{\text{B}})\cos(\theta_{\text{A}} + \alpha) \nonumber\\
 &\hphantom{{}=I_0(\beta M r_{\text{B}})\inv\exp\Big\{}+ K_2(m_{\text{A}} - m_{\text{B}})\sin(\theta_{\text{A}} + \alpha)] \Big\} 
 \label{eq:rho_A} \\
 \rho_{\text{B}} &= I_0(\beta M r_{\text{A}})\inv\exp\Big\{\beta r_{\text{A}}[(1 + K_1 m_{\text{A}} m_{\text{B}})\cos(\theta_{\text{B}} - \alpha) \nonumber\\
 &\hphantom{{}=I_0(\beta M r_{\text{A}})\inv \exp\Big\{}+ K_2(m_{\text{A}} - m_{\text{B}})\sin(\theta_{\text{B}} - \alpha)] \Big\}
 \label{eq:rho_B}
\end{align}

Using \cref{eq:rho_A,eq:rho_B} in \cref{eq:f_E,eq:f_S}, performing the integrals, and rescaling the free energy density, \cref{eq:f_general}, by $f \gets z\inv f$, leads to 
\begin{widetext}
 \begin{multline}
  2 f = - \big[(1 + K_1 m_{\text{A}} m_{\text{B}})\cos(2\alpha) + K_2(m_\text{A} - m_\text{B})\sin(2\alpha)\big]R_1(M,r_{\text{A}})R_1(M,r_{\text{B}}) \\
  + M r_\text{A} R_1(M,r_{\text{A}})
  + M r_\text{B} R_1(M,r_{\text{B}})
  + \beta\inv \Big[s_{\groupZ}(m_\text{A}) + s_{\groupZ}(m_\text{B}) - \ln\big(I_0(\beta M r_\text{A})\big) - \ln\big(I_0(\beta M r_\text{B})\big)\Big].
 \label{eq:f_intermediate}
 \end{multline}
\end{widetext}
Here, we have introduced the shorthand notation
\begin{equation}
 R_l(M,r) \equiv \frac{I_l(\beta M r)}{I_0(\beta M r)}.
\end{equation}
\Cref{eq:f_intermediate} is to be minimized over $m_\text{A}$, $m_\text{B}$, $r_\text{A}$, $r_\text{B}$, and $\alpha$.

Due to symmetry, $r_\text{A} = r_\text{B} \equiv r$ and $s_{\groupZ}(m_\text{A}) = s_{\groupZ}(m_\text{B})$. By differentiating \cref{eq:f_intermediate} with respect to $\alpha$ we find the minimizing condition
\begin{equation}
 \tan(2\alpha) = \frac{K_2(m_\text{A} - m_\text{B})}{1 + K_1 m_{\text{A}} m_{\text{B}}},
 \label{eq:canting_angle}
\end{equation}
or
\begin{align}
 \cos(2\alpha) &= \frac{1 + K_1 m_{\text{A}} m_{\text{B}}}{M} \\
 \sin(2\alpha) &= \frac{K_2(m_\text{A} - m_\text{B})}{M}.
\end{align}
Using these facts, \cref{eq:f_intermediate} can be simplified to
\begin{multline}
 f = M \Big[ r R_1 - \tfrac{1}{2}{R_1}^2 \Big] \\
  + \beta\inv \Big[s_{\groupZ}(m) - \ln\big(I_0(\beta M r)\big) \Big].
  \label{eq:f_final}
\end{multline} 
In the continuation, we have to distinguish between the \enquote{ferromagnetic sector}, where $m_{\text{A}} = m_{\text{B}}$, and the \enquote{antiferromagnetic sector}, where $m_{\text{A}} = - m_{\text{B}}$. In the two sectors we have
\begin{align}
 M_{\text{fm}} &= 1 + K_1 m_\text{fm}^2, \label{eq:M_fm}\\
 M_{\text{afm}} &= \sqrt{(1 - K_1 m_\text{afm}^2)^2 + 4{K_2}^2 m_\text{afm}^2}, \label{eq:M_afm}
\end{align}
with
\begin{align}
 m_\text{fm} &\equiv \tfrac{1}{2}(m_\text{A} + m_\text{B}), \label{eq:m_fm}\\
 m_\text{afm} &\equiv \tfrac{1}{2}(m_\text{A} - m_\text{B}). \label{eq:m_afm}
\end{align}
as order parameters. Since
\begin{equation}
 s_{\groupZ} (m_\text{fm}) = s_{\groupZ} (m_\text{afm})
\end{equation}
we may drop the subscript \enquote{(a)fm} and just write $M$ and $m$ from now on, as long as we are cautious of which version, \cref{eq:M_fm,eq:m_fm} or \cref{eq:M_afm,eq:m_afm}, to apply.

\subsection{Determining the mean field phase diagram}
The remaining task in obtaining the phase diagram is basically to minimize \cref{eq:f_final}. First we note that if the system is $\groupU$ symmetric, hence $r=0$, \cref{eq:f_final} reads
\begin{equation}
 f = \beta\inv s_{\groupZ}(m),
\end{equation}
which has a global minimum at $m=0$ for $\beta\inv>0$. In other words: There are \emph{no} $\groupZ$ broken ($m>0$) $\groupU$ symmetric ($r=0$) mean field solutions of the model.

On the other hand, if $m=0$ (corresponding to $M = 1$) the free energy is that of an ordinary $XY$-model,
\begin{equation}
 f_\text{XY} = rR_1(1,r) - \tfrac{1}{2}R_1(1,r)^2 - \beta\inv\ln\big(I_0(\beta M r)\big).
 \label{eq:f_XY}
\end{equation}
which displays a second order phase transition at $\beta_{\groupU}=2$.

We will now assume that $r = r_\text{XY}$, where $r_\text{XY}$ is the $r$ minimizing $f_\text{XY}$, and Taylor expand the free energy density about $m=0$ to determine the nature of the $\groupZ$ transition. $r_\text{XY}$ is plotted in \cref{fig:r_XY}.
\begin{figure}[]
\includegraphics{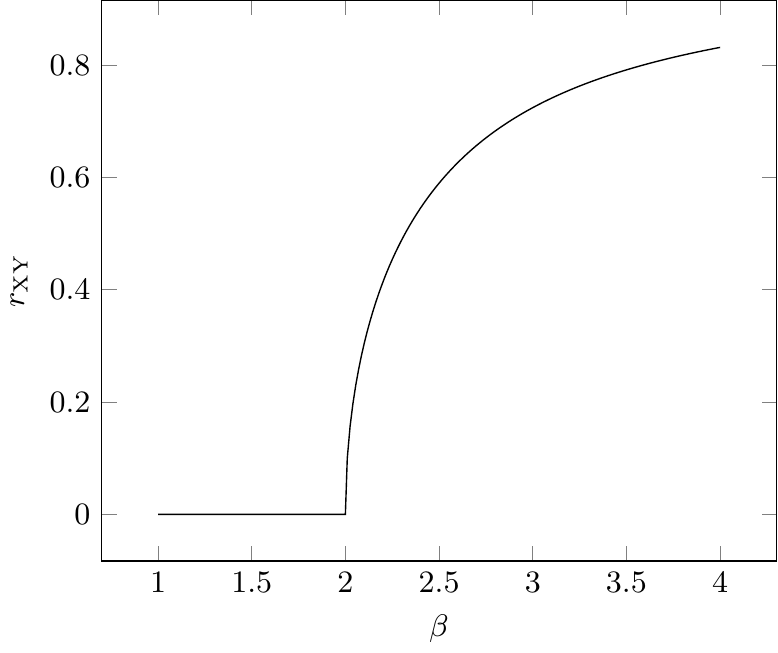}
\caption{$r_{\text{XY}}(\beta)$, the $r(\beta)$-value minimizing $f_\text{XY}(\beta)$, \cref{eq:f_XY}.}
\label{fig:r_XY}
\end{figure}
It is legitimate to put $r = r_\text{XY}$ when approaching the $\groupZ$ transition, as the free energy density, \cref{eq:f_final}, is analytic. We write
\begin{equation}
 f(m) \approx f(0) + \frac{a_2}{2}m^2 + \frac{a_4}{4}m^4 + \frac{a_6}{6}m^6 + \cdots
 \label{eq:f_taylor}
\end{equation}
If $a_2 > 0, a_4 > 0$ there is a single minimum at $m=0$, i.e. no symmetry breaking. If $a_2 = 0, a_4 > 0$ we have second order phase transition to a symmetry broken state, whereas we have a first order transition if $a_2>0,a_4 = -4\sqrt{a_2a_6}/\sqrt{3} < 0$. Tricriticality is achieved when $a_2 = a_4 = 0$. Expanding \cref{eq:f_final} gives
\begin{widetext}
\begin{align}
 a_{2,\text{fm}} &=  -\big[(2{R_2}^2 - R_2 - 1)(r^2 - R_1r)\beta + {R_1}^2\big]K_1 + \beta\inv, \label{eq:a_2_fm}\\
 a_{2,\text{afm}} &= -\big[(2{R_2}^2 - R_2 - 1)(r^2 - R_1r)\beta + {R_1}^2\big]\big[2{K_2}^2 - K_1\big] + \beta\inv, \label{eq:a_2_afm}\\
 \begin{split}
  a_{4,\text{fm}} &= \tfrac{1}{2}\big[(8{R_1}^3 - 6 R_1 R_2 - 3R_1 + R_3)\beta^2r^3 - (12{R_1}^4 - 10{R_1}^2R_2 - 7{R_1}^2 + {R_2}^2 + R_1R_3 + 2R_2 + 1)\beta^2r^2 \\
  & \hphantom{{}= \tfrac{1}{2}\big[} + 2(2{R_1}^2 - R_2 -1)(2R_1 - r)\beta r\big]{K_1}^2 + \tfrac{1}{3}\beta\inv ,
 \end{split} \label{eq:a_4_fm}\\
 \begin{split}
  a_{4,\text{afm}} &= \tfrac{1}{2}\big[(8{R_1}^3 - 6 R_1 R_2 - 3R_1 + R_3)\beta^2r^3 - (12{R_1}^4 - 10{R_1}^2R_2 - 7{R_1}^2 + {R_2}^2 + R_1R_3 + 2R_2 + 1)\beta^2r^2 \\
  & \hphantom{{}= \tfrac{1}{2}\big[} + 2R_1(2{R_1}^2 - R_2 -1)\beta r + 2{R_1}^2\big]\big[2{K_2}^2 - K_1\big]^2 + \big[(2{R_1}^2 - R_2 -1)(R_1 - r)\beta r - {R_1}^2\big]{K_1}^2 + \tfrac{1}{3}\beta\inv .
 \end{split} \label{eq:a_4_afm}
\end{align} 
\end{widetext}
Now $R_n$ is shorthand notation for $R_n(1,r)$.

First we observe, from \cref{eq:a_2_fm,eq:a_2_afm} and the expansion \eqref{eq:f_taylor}, that as long as $K_1 > 2{K_2}^2 - K_1$, or $K_1 > {K_2}^2$, and $r$ is finite, $f_{\text{fm}} < f_\text{afm}$ and the $\groupZ$ ordering will be ferromagnetic. $K_1 = {K_2}^2$ determines the border between the two sectors, where there there can be no $\groupZ$ ordering. Mathematically this reasoning holds only as long as we are expanding around $m=0$. Could there be a ferromagnetic-antiferromagnetic transition somewhere within the $\groupZ$ ordered phase, i.e. for larger $m$ values where the Taylor expansion breaks down? The answer is \emph{no}, because the Taylor expansion never breaks down near a $\groupZ$ transition, be it a disorder-order or ferromagnetic-antiferromagnetic transition. This follows from the fact that in the ferromagnetic phase $m_\text{fm} > 0$, but $m_\text{afm} = 0$, and vice versa (see the definitions, \cref{eq:m_fm,eq:m_afm}).

By numerical minimization of \cref{eq:f_final} we find that the $\groupZ$ transition is second order for sufficiently small $(K_1,K_2)$ values and first order for sufficiently large $(K_1,K_2)$ values. It is always separated by a finite temperature interval from the $\groupU$ transition. The tricritical line (in the $\groupZ$ sector) in the $(K_1,K_2)$ plane is then found by solving $a_2 = 0, a_4 = 0$, both in the ferromagnetic and the antiferromagnetic sector.

The final standard mean-field result is shown in \cref{fig:MF_phasediagram}.

\section{Cluster mean field calculations}\label{app:CMF}
The mean field theory may be refined by extending the number of lattice sites decoupled from neighboring sites by means of a mean field, from one to a cluster of several.~\cite{Oguchi,Yamamoto} In this way we may capture some of the fluctuation effects that are supressed in the standard single-site mean field calculations, while keeping the results (numerically) exact. These cluster mean field (CMF) results, as leading order corrections to the mean field theory, provide an indication of how strong the fluctuation effects are.

In this work we have focused on a cluster of $2\times 2\times 2$ sites, at the border coupled to the mean fields $m$ and $\v r$ in the $\groupZ$ and $\groupU$ sector, respectively. We label them A and B depending on which sublattice they belong to. The internal links in the cluster are treated exactly, and hence they do not have to be labeled (apart from their coordinates).

The CMF Hamiltonian reads
\begin{widetext}
 \begin{align}
  H_{\text{CMF}} &= -\sum_{\langle i,j \rangle} (1 + K_1 \sigma_i \sigma_j)\cos(\theta_i - \theta_j) + K_2 (\sigma_i - \sigma_j)\sin(\theta_i - \theta_j) \\
  &\hphantom{={}} - \frac{z}{2}\sum_{i \in \text{A}} (1 + K_1 \sigma_i m_B)r \cos(\theta_i + \alpha) + K_2(\sigma_i - m_B) r\sin(\theta_i + \alpha) \\
  &\hphantom{={}} - \frac{z}{2}\sum_{i \in \text{B}} (1 + K_1 \sigma_i m_A)r \cos(\theta_i - \alpha) + K_2(\sigma_i - m_A) r\sin(\theta_i - \alpha).
 \end{align}
\end{widetext}
The $\frac{z}{2}$ prefactor comes from the fact that half of the neighboring sites of a given site in the cluster is \enquote{mean field approximated} sites outside the cluster. This factor will in general, for other cluster shapes and sizes than $2 \times 2 \times 2$, be site dependent. $\alpha$ is half of the canting angle between $\v r_\text{A}$ and $\v r_\text{B}$, as in the MF calculations above. As a first approximation we assume it to be given by the MF expression, \cref{eq:canting_angle}, with $m_\text{A(B)}$ as determined in the self-consistent CMF calculation.

In order to obtain the partition function $\mathcal{Z}_\text{CMF} = \int \exp(-\beta H_\text{CMF})$ we have to integrate out the degrees of freedom associated with the cluster. Performing the sum over all $\sigma$ configurations (just $2^8 = 256$ terms) is easily done on a computer. A closed form integral of the $\theta$ configurations is however not known to the authors, but by mapping the partition function to a \enquote{link current} model~\cite{PhysRevB.88.094412,PhysRevLett.87.160601} we may obtain a convergent series of the weights associated with the \enquote{current} configurations, which, given an appropriate cutoff, can also be handled with a computer. The basic idea is to write the cosines and sines on their complex forms, $\cos x = (\e{\i x} + \e{-\i x})/2$ and $\sin x = (\e{\i x} - \e{-\i x})/(2\i)$, Taylor expand all the terms of the Boltzmann factor $\exp(-\beta H_\text{CMF})$ (a so-called \enquote{high temperature expansion}), separate out the $\e{\i\theta_i f_i}$ terms for each $i$ (which is now possible), and then perform the $\theta_i$-integrals, each leading to a Kronecker $\delta$-function forcing the constraint $f_i=0$. Here, $f_i$ is a function of the Taylor expansion coefficients associated with site $i$, which can be interpreted as the sum of integer currents flowing into site $i$ along the connecting lattice links. After some algebra, relabelling and identification of the series expansions of modified Bessel functions, we end up with
\begin{widetext}
 \begin{align}
  \mathcal{Z}_\text{CMF} &= \sum_\sigma \sum_{\nabla (k + l) = 0} \prod_{\langle i,j\rangle} I_{|k_{ij}|}(\beta X_{ij})\e{\i k_{ij} x_{ij}} \prod_{i \in \text{A}} I_{|l_i|}(\tfrac{z}{2}\beta r Y_{i\text{A}})\e{\i\l_i y_{i\text{A}}}\prod_{i \in \text{B}} I_{|l_i|}(\tfrac{z}{2}\beta r Y_{i\text{B}})\e{\i\l_i y_{i\text{B}}}, \label{eq:Z_CMF}\\
  X_{ij} &= \sqrt{1 + {K_1}^2 + 2{K_2}^2 + 2(K_1 - {K_2}^2)\sigma_i\sigma_j}, \\
  x_{ij} &= \arctan\left(\frac{K_2(\sigma_i - \sigma_j)}{1 + K_1 \sigma_i\sigma_j}\right), \\
  Y_{i\text{A(B)}} &= \sqrt{1 + {K_1}^2{m_{\text{B(A)}}}^2 + {K_2}^2(1 + {m_{\text{B(A)}}}^2) + 2(K_1 - {K_2}^2)\sigma_i m_{\text{B(A)}}}, \\
  y_{i\text{A(B)}} &= \arctan\left(\frac{K_2(\sigma_i - m_{\text{B(A)}})}{1 + K_1 \sigma_im_{\text{B(A)}}}\right) + (-) \alpha.
 \end{align}
\end{widetext}
$k_{ij} \in \mathbb{Z}$ denote a current along the link from sites $i$ to $j$ \emph{within} the cluster, while $l_i \in \mathbb{Z}$ is a current \emph{leaving} the cluster from site $i$. The $\nabla (k + l) = 0$ subscript of the summation means that only configurations where current conservation is enforced for all sites (because of the $f_i=0$ constraints) are included.

In the ferromagnetic sector we write $m = m_\text{A} = m_\text{B}$ and in the antiferromagnetic $m = m_\text{A} = -m_\text{B}$. $m$ is (up to an arbitrary sign) given by
\begin{equation}
 m = \mathcal{Z}_\text{CMF}\inv[\mathcal{Z}'_\text{CMF} (\sigma_1=1) - \mathcal{Z}'_\text{CMF}(\sigma_1=-1) ]
 \label{eq:m_selfcons}
\end{equation}
where $\mathcal{Z}'_\text{CMF}$ is the \enquote{reduced} partition function where we have kept one $\sigma$ spin fixed (here: $\sigma_1$) when integrating out the degrees of freedom. $r$ is found by differentiation of the partition function:
\begin{equation}
 r = \mathcal{Z}_\text{CMF}\inv \pderiv{\mathcal{Z}_\text{CMF}}{(\tfrac{z}{2}\beta Y_{1})}
 \label{eq:r_selfcons}
\end{equation}

We find $m = m(\beta,K_1,K_2)$ and $r = r(\beta,K_1,K_2)$ -- and by this the CMF phase diagram of the model -- by self-consistently solving the coupled nonlinear \cref{eq:m_selfcons,eq:r_selfcons}. This is done numerically. To make this a tractable task we have to impose an upper cutoff, $k_\text{max}$, on the allowed magnitude of $|k_{ij}|$  (by current conservation $\set{l_i}$ is given once $\set{k_{ij}}$ is known). Fortunately, the terms in the partition function, \cref{eq:Z_CMF}, are highly convergent in $k_\text{max}$. In our calculations we used $k_\text{max} = 3$, which was found to be sufficient to yield correct CMF transition temperatures to about 11 significant digits (found by comparing with single $k_\text{max} = 4$ and $k_\text{max} = 5$ calculations.)

The final cluster mean-field result is shown in \cref{fig:CMF_phasediagram}.

\section{Monte Carlo simulations}\label{app:MC}
In this work, the Monte-Carlo method of choice has been Wang--Landau (WL) sampling.~\cite{PhysRevLett.86.2050,PhysRevE.64.056101} The main motivation for this is that broad histogram methods, like the WL algorithm, compares favorably to ordinary canonical sampling in dealing with models having rough energy landscapes (caused by frustration in this case) and (possible) first order phase transitions. Furthermore, the broad range of energies traversed in one WL simulation means that the properties of the model may be determined in a single run, as opposed to a canonical simulation where, if the temperatures of interest are not known a priori, separate computations for a range of temperatures must be performed. This is of practical, labor saving significance when exploring the large parameter space of $(\beta,K_1,K_2)$.

For a more complete discussion and details on the procedure we refer to the Appendices E and F of Ref. \onlinecite{PhysRevB.89.104509}.

\bibliography{references}

\end{document}